\documentclass[10pt,a4paper]{amsart}

\usepackage{amsthm,amssymb,amsmath,epic,eepic,float}
\usepackage{rotating,epsfig,array,varioref}
\usepackage[shortcuts]{extdash}

\usepackage{color}

\def\bb{\bar{b}}
\def\tkappa{\tilde{\kappa}}

\def\ll{\left\lgroup}
\def\rr{\right\rgroup}

\def\leq{\leqslant}
\def\geq{\geqslant}

\def\zi{Z_{L \times L}}
\def\zs{Z_{N \times N}}

\newtheorem{myLemma}{Lemma}

\restylefloat{figure}

\newcommand{\cH}{\mathcal{H}}

\newcommand{\cN}{\mathcal{N}}

\def\ll{ \left\lgroup}
\def\rr{\right\rgroup}

\newcommand{\1}{{\bf 1.}}
\newcommand{\2}{{\bf 2.}}
\newcommand{\3}{{\bf 3.}}
\newcommand{\4}{{\bf 4.}}
\newcommand{\5}{{\bf 5.}}
\newcommand{\6}{{\bf 6.}}
\newcommand{\7}{{\bf 7.}}
\newcommand{\8}{{\bf 8.}}

\def\det{\operatorname{det}}

\begin{document}

\title[Variations on Slavnov's scalar product]%[Slavnov's scalar products]
{Variations on Slavnov's scalar product}

\author[O Foda and M Wheeler]
{O Foda $^1$ and M Wheeler $^2$}

\address{
\!\!\!\!\!\!\!$^1$ Dept of Mathematics and Statistics,
University of Melbourne,
Parkville, VIC 3010, Australia
\newline
$^2$ Laboratoire de Physique Th\'eorique et Hautes Energies,
CNRS UMR 7589 and Universit\'e Pierre et Marie Curie - Paris 6,
4 place Jussieu, 75252 Paris cedex 05, France
}

\email{omar.foda@unimelb.edu.au, mwheeler@lpthe.jussieu.fr}


\begin{abstract}
We consider the rational six-vertex model on an $L\! \times \! L$ 
lattice with domain wall boundary conditions and restrict $N$ 
parallel-line rapidities, $N \leq L/2$, to satisfy length-$L$ 
XXX spin-$\frac{1}{2}$ chain Bethe equations. 
We show that the partition function is an $(L-2N)$-parameter 
extension of Slavnov's scalar product of a Bethe eigenstate 
and a generic state, with $N$ magnons each, on a length-$L$ 
XXX spin-$\frac{1}{2}$ chain. 
Decoupling the extra parameters, we obtain a third determinant 
expression for the scalar product, where the first is due to 
Slavnov \cite{slavnov}, and the second is due to 
Kostov and Matsuo \cite{kostov.matsuo}.
We show that the new determinant is Casoratian, and consequently 
that tree-level $\cN \! = \! 4$ SYM structure constants that are 
known to be determinants, remain determinants at 1-loop level. 
\end{abstract}

\maketitle
\setcounter{section}{-1}

\section{Introduction}
\label{section.Introduction}

Scalar products of $N$-magnon states on a length-$L$ spin chain, 
play a central role in studies of correlation functions in 
integrable spin chains \cite{korepin.book, kmt}. Recently, they 
have appeared in studies of 3-point functions in 4-dimensional 
$\cN \! = \! 4$ super Yang-Mills theory, SYM$_4$ 
\cite{E1, E2, GSV, Foda}. 
Of particular interest is the scalar product of an eigenstate 
of the spin-chain transfer matrix and a generic state, which 
in the case of integrable XXX and XXZ spin-$\frac{1}{2}$ chains 
was evaluated by N Slavnov as an $N\! \times \! N$ determinant 
\cite{slavnov}. 
Recently, I Kostov and Y Matsuo obtained a second expression 
for the same object as a $2N\! \times \! 2N$ 
determinant \cite{kostov.matsuo}. 

In this work, we start from Izergin's determinant expression 
for the domain wall partition function 
of the rational six-vertex model on an $L\! \times \! L$ 
lattice \cite{izergin}, and require that the rapidities on 
$N$ parallel
lattice lines, $N \leq L/2$, satisfy the Bethe equations 
of a length-$L$ XXX spin-$\frac{1}{2}$ chain. We show that 
the result is an extended version of Slavnov's scalar 
product that depends on $(L - 2N)$ extra parameters. 
Taking these extra parameters to infinity, so they decouple 
from the partition function, we obtain a third expression 
for Slavnov's scalar product as an $L\! \times \! L$ determinant. 
We show that the new determinant expression is a discrete KP 
$\tau$-function in the inhomogeneities that can be written 
in Casoratian form (the discrete analogue of a Wronskian). 
This allows to use the results of N Gromov and P Vieira
\cite{gromov.vieira.short, gromov.vieira.long} 
to prove that SYM$_4$ tree-level structure constants that 
are known to be determinants \cite{Foda}, remain determinants 
at 1-loop level.

\subsection{Outline of contents}
In Section
   {\bf 1}, we recall basic definitions to make the presentation 
reasonably self-contained, and review recent results to put our
own results in context. 
In {\bf 2}, we recall basic facts related to the scalar product. 
In {\bf 3}, we require that a subset of the parameters of 
Izergin's domain wall partition function are Bethe roots, 
and identify the result as a scalar product with extra
parameters.
In {\bf 4}, we decouple the extra parameters of Section {\bf 3}, 
to obtain a third determinant expression of Slavnov's scalar. 
In {\bf 5}, we prove that the third determinant expression is 
equal to the second determinant expression of \cite{kostov.matsuo}.
In {\bf 6}, we give a new proof that the second determinant 
expression of Kostov and Matsuo is equal to the first determinant 
expression of Slavnov. Our proof is along the lines of Izergin's 
proof of the determinant expression of the domain wall partition 
function, and can be regarded as an alternative to the proof in 
\cite{kostov.matsuo}.
In {\bf 7}, we show that the structure constants that were expressed 
in determinant form in \cite{Foda}, retain their determinant form 
when 1-loop radiative corrections are included along the lines of 
\cite{gromov.vieira.short, gromov.vieira.long}. 
Finally, in {\bf 8}, we collect a number of remarks.

\section{Definitions and overview}
\label{section.Overview}

\subsection{Context, notation, {\it etc.} used in this work}
We restrict our attention to 
the rational      six-vertex model and 
XXX spin-$\frac{1}{2}$ chain, 
but our conclusions extend to 
the trigonometric six-vertex model and 
XXZ spin-$\frac{1}{2}$ chain, as well as to vertex models 
based on higher-spin $su(2)$ representations and 
spin chains. 
All six-vertex configurations 
will have $L$ vertical lattice lines, and all spin chains 
will be of length $L$ and periodic. 

In six-vertex terms, 
$\{x\}$ is a set of free rapidities that flow in 
horizontal lattice lines, 
$\{y\}$ is a set of free rapidities that flow in 
vertical lattice lines, and 
$\{b\}$ is a set of rapidities that flow in 
horizontal lattice lines and satisfy 
length-$L$ spin-$\frac{1}{2}$ chain Bethe equations. 
From now on, we refer 
to the rapidities that flow in horizontal lattice lines as 
\lq rapidities\rq, and 
to the rapidities that flow in vertical   lattice lines as 
\lq inhomogeneities\rq.

In spin-chain terms,
$\{x\}$ is a set of free auxiliary space rapidities,
$\{y\}$ is a set of free quantum   space rapidities, 
or inhomogeneities, and 
$\{b\}$ is a set of auxiliary space rapidities that 
satisfy Bethe equations. 
From now on, and similarly to the six-vertex case, we refer 
to the auxiliary space rapidities as 
\lq rapidities\rq, 
and to the quantum space rapidities as 
\lq inhomogeneities\rq.
We use $|z|$ for the cardinality of a set $\{z\}$. 

Rapidities that satisfy Bethe equations are referred to as 
\lq Bethe-restricted\rq. Six-vertex model configurations
and partition functions that depend on Bethe-restricted 
variables are also Bethe-restricted. Partitions functions 
with a subset of rapidities set equal to a subset of the 
inhomogeneities are \lq inhomogeneity\-/restricted\rq.  

\subsection{Bethe eigenstates and a generic states in XXX 
spin-$\frac{1}{2}$ chains}
Consider a length-$L$ periodic integrable XXX spin-$\frac{1}{2}$ 
chain. The Hilbert space of states $\cH$ is spanned by magnon 
states.
An $N$-magnon state, $N = 0, 1, 2, \dots$, is created by 
the action of $N$ Bethe raising-operators $B(x_i)$, 
where $x_i$, $i = 1, 2, \dots, N$, are free rapidities, on a 
pseudo-vacuum state $| \textit{vac} \rangle$.
A dual Hilbert space of states $\cH^{\star}$ is analogously 
created by the action of Bethe lowering-operators $C(x_i)$ 
on a dual pseudo-vacuum state $\langle \textit{vac} |$.
For more details using the same notation and terms used 
in this work, see \cite{foda.wheeler.jimbo.fest}.

States characterized by free 
rapidities $\{x\}$  
are not eigenstates of the spin chain transfer matrix. 
They are \lq generic\rq,           or \lq off-shell\rq. 
States characterized by Bethe-restricted 
rapidities $\{b\}$ 
are     eigenstates of the spin chain transfer matrix. 
They are \lq Bethe eigenstates\rq, or \lq on-shell\rq. 

\subsection{The scalar product of a Bethe eigenstate and 
a generic state}
Scalar products of two magnon states play an essential role 
in studies of integrable spin chains. 
If both states are off-shell, 
$|x_1 \rangle$ and 
$|x_2 \rangle$, then the scalar product 
$\langle x_1 | x_2 \rangle$ 
$=$
$\langle x_2 | x_1 \rangle$ can be expressed in Izergin-Korepin 
sum form \cite{korepin.book}.
If both states are on-shell, 
$|b_1 \rangle$ and 
$|b_2 \rangle$, then the scalar product vanishes unless
$\{b_1\} = \{b_2\} = \{b\}$. In that case, 
$\langle b | b \rangle$ 
is the Gaudin norm \cite{gaudin.book, korepin.paper}. 

If one state is on-shell, 
$| b \rangle$,   
and the other off-shell, 
$| x \rangle$, 
then the scalar product 
$\langle b | x \rangle$ = 
$\langle x | b \rangle$ 
can be evaluated in determinant form. 
This is the case in which we are primarily interested in this work.

\subsection{The first determinant expression for the scalar product}

For spin chains with $su(2)$-symmetry, such as the XXX and XXZ 
spin-$\frac{1}{2}$ chains, and their higher-spin analogues, 
$\langle b | x \rangle$ was evaluated by Slavnov 
in determinant form \cite{slavnov}, 
and therefore is frequently referred to as Slavnov's scalar product. 
In this work, we simply say 
\lq scalar product\rq, and, in light of the results in 
\cite{kostov.matsuo} and in this work, we refer to Slavnov's 
determinant expression as 
\lq the first determinant expression for the scalar product\rq.
No tractable expression, such as a determinant, is 
known for scalar products of an off-shell state and an on-shell 
state in integrable models based on higher rank algebras. 

\subsection{The scalar product is a discrete KP $\tau$-function} 

The scalar product $\langle b| x \rangle$ of two $N$-magnon 
states is a function of three sets of variables: 
\1 A set of Bethe-restricted rapidities 
$\{b\}$, of cardinality $N$, 
\2 A set of free             rapidities 
$\{x\}$, of cardinality $N$, 
\3 A set of free             inhomogeneities 
$\{y\}$, of cardinality $L$, 
where $L$ is the number of sites of the spin chain that the states 
live on. To simplify the notation, we will frequently omit to show 
the dependence on $\{y\}$. 

In \cite{foda.wheeler.zuparic, foda.schrader}, it was shown that 
$\langle b| x \rangle$ is a discrete KP $\tau$-function
in $\{x\}$. Since $\langle b| x \rangle$ is symmetric in 
$\{y\}$ as well, it was conceivable that $\langle b| x \rangle$ 
is a $\tau$-function in $\{y\}$ as well, but it was not 
straightforward to show that. 

\subsection{Inhomogeneity\-/restricted scalar products}

In \cite{kmt, wheeler.su2}, a class of restricted scalar products, 
obtained by setting a subset of $\{x\}$ equal to a subset of 
$\{y\}$, was studied. Since the scalar products are determinants, 
the inhomogeneity\-/restricted scalar products are also determinants. 

In \cite{wheeler.su2}, the six-vertex model configurations whose 
partition functions are scalar products, and their step-by-step 
restrictions that lead to inhomogeneity\-/restricted scalar products 
were studied in detail. 

\subsection{Tree-level SYM$_4$ structure constants that can be 
expressed in determinant form}

In \cite{E1, E2, GSV}, a class of tree-level 3-point functions 
of states that live in scalar $su(2)$ 
sectors\footnote{More than one $su(2)$ sector is involved in 
these 3-point functions.} were formulated in XXX spin-$\frac{1}{2}$ 
chain terms, and Izergin-Korepin sum expressions were obtained for 
their structure constants.  

In \cite{Foda}, the tree-level structure constants of 
\cite{E1, E2, GSV} were identified with the 
inhomogeneity\-/restricted scalar products of 
\cite{kmt, wheeler.su2}.  The basic idea of \cite{Foda} was to 
formulate the structure constants in six-vertex terms, compare 
the result with the analogous formulation of the 
inhomogeneity\-/restricted scalar products in six-vertex terms of 
\cite{wheeler.su2}, and 
to show that the two objects are equal, up to an overall factor 
that is easily computed.

In \cite{GSV}, a special case of the 3-point functions of \cite{E1, E2} 
where one operator is a non-BPS, while the other two operators are 
(essentially) BPS states was studied in detail. It turns out that 
these objects can also be expressed in terms of determinants that 
are obtained from inhomogeneity\-/restricted Slavnov determinants by 
taking a set of Bethe roots to infinity. 

\subsection{The \lq theta morphism\rq\ of Gromov and Vieira}
In studies of quantum integrability in weakly-coupled $\cN = 4$ 
supersymmetric Yang-Mills, SYM$_4$, gauge-invariant single-trace 
composite operators are mapped to states in closed spin chains
\cite{beisert.review}. In particular, tree-level single-trace 
operators, in $su(2)$ scalar sectors, that are eigenstates of 
the 1-loop mixing matrix, are mapped to states of periodic XXX 
spin-$\frac{1}{2}$ chains. 
So far, these spin chains were homogeneous in the sense that 
the inhomogeneities were set to the same value, which can be 
set to zero.

In \cite{gromov.vieira.short, gromov.vieira.long}, N Gromov and 
P Vieira showed that 1-loop radiative corrections can be introduced 
into the structure constants studied in \cite{E1, E2, GSV}, and 
that were expressed as determinants in \cite{Foda}, by switching 
on the inhomogeneities, that is, by choosing 
$y_i = \theta$, $i = 1, \dots, L$, computing 
the structure constants to lowest non-trivial order in $\theta$, 
which is $O(\theta^2)$, as there are no order $O(\theta)$ 
contributions, and setting $\theta$ equal to the gauge coupling 
constant.

\subsection{1-loop     SYM$_4$ structure constants that can be
expressed in determinant form}

In \cite{foda.wheeler.partial}, structure constants 
of the latter 3-point functions were identified with six-vertex model 
configurations on $(N \times L)$-rectangular lattices, $N \leq L$, with 
\lq partial domain wall boundary conditions\rq, and the corresponding 
\lq partial domain wall partition functions\rq, pDWPF's, were studied 
in some detail. 

In \cite{foda.wheeler.partial}, 
we obtained two expressions for these partition functions: 
\1 As a determinant of an $L\times L$ matrix, $\zi$ and 
\2 As a determinant of an $N\times N$ matrix, $\zs$. 
The latter was first obtained by I Kostov 
\cite{kostov.short.paper, kostov.long.paper}.
Further, we showed that 
$\zi$ is a discrete KP $\tau$-function in the inhomogeneities $\{y\}$, 
and  
$\zs$ is a discrete KP $\tau$-function in the free rapidities $\{x\}$.

Using the fact that these pDWPF's are discrete KP 
$\tau$-functions in the inhomogeneities $\{y\}$, together with 
the recent results of N Gromov and P Vieira 
\cite{gromov.vieira.short, gromov.vieira.long} allowed us to show 
that these structure constants remain determinants in the presence 
of 1-loop radiative corrections. 
On the other hand, the fact that we were unable to show that the 
inhomogeneity\-/restricted scalar product is a $\tau$-function in the 
inhomogeneities 
prevented us from extending the determinant result to 1-loop level
for tree-level structure constants with 3 non-BPS operators. 

\subsection{A second determinant expression for the scalar product}

In \cite{kostov.private.communication.e.mail, 
        smirnov.private.communication},
I Kostov and F Smirnov independently suggested that the scalar product 
of a Bethe eigenstate and a generic state can be obtained from Izergin's 
$(L \times L)$ domain wall partition function, either by sending 
$(L - 2N)$ rapidities, $N \leq L/2$, to infinity, and thereby 
decoupling them so that one ends up with a partial domain wall 
partition function, then setting $N$ of the remaining 
rapidities to satisfy appropriate Bethe equations 
\cite{kostov.private.communication.e.mail}, or by re-interpreting 
Korepin's domain wall configuration as the scalar product of a Bethe 
eigenstate that is built on the lowest-weight pseudo-vacuum (all spins 
down, rather than up) and a generic state, by requiring an appropriate 
subset of rapidities to satisfy appropriate Bethe equations
\cite{smirnov.private.communication}. The latter Bethe equations
should come from \lq beyond the equator\rq\ \cite{pronko.stroganov}.
 
In \cite{kostov.matsuo}, I Kostov and Y Matsuo obtained a realization 
of the suggestion of \cite{kostov.private.communication.e.mail}, 
starting from $2N \times L$ lattice configurations, $2N \leq L$, 
called \lq partial domain wall configurations\rq\ 
in \cite{foda.wheeler.partial}, whose partition functions can be 
written in determinant form in two different ways:
As the determinant of an $(2N\times 2N)$ matrix 
\cite{kostov.short.paper, kostov.long.paper}, or as the determinant 
of an $(L\times L)$ matrix. In \cite{kostov.matsuo}, Kostov and Matsuo 
start from the $(2N \times 2N)$ determinant, require that $N$ rapidities 
satisfy an appropriate set of Bethe equations, then show that the result 
is equal to Slavnov's $(N\times N)$ determinant expression for the scalar 
product.

\subsection{Bethe-restricted domain wall partition functions as 
parameter-ex\-t\-ended scalar products}
In this work, we obtain a parameter-extension of 
\cite{kostov.private.communication.e.mail,
kostov.matsuo}, that is not identical to but partially along the lines 
of \cite{smirnov.private.communication}. 
We start from an $(L\times L)$ Korepin domain wall configuration, 
and the corresponding  $(L\times L)$ Izergin determinant expression for 
the partition function, then
we set $N$ rapidities to satisfy appropriate Bethe 
equations, such that Izergin's determinant now has $(L-N)$ free 
rapidities, and split the latter into two subsets
$\{x\}$ of cardinality $N$, and 
$\{t\}$ of cardinality $(L - 2N) \geq 0$.  
We interpret the resulting determinant as an $(L - 2N)$-parameter 
extension of the scalar product of an $N$-magnon Bethe eigenstate
and an $N$-magnon generic state, on a length-$L$ XXX spin-$\frac{1}{2}$ 
chain. Everything we say applies without obstruction to six-vertex models 
with trigonometric weights and the corresponding XXZ spin-$\frac{1}{2}$  
chains. 

\subsection{A third determinant expression for the scalar product}
Taking the $(L - 2N)$-extension parameters to infinity, having first 
normalized properly, we obtain a determinant of an $(L\times L)$ 
matrix, that we interpret as a third determinant expression of 
the scalar product of a Bethe eigenstate and a generic state, 
of $N$ magnons each, on a length-$L$ XXX spin-$\frac{1}{2}$ chain.

We prove that by showing that our determinant expression is equal
to the second determinant expression in \cite{kostov.matsuo}.
The fact that the two have different forms follows from the 
fact that \cite{kostov.matsuo} start from the $(2N\times 2N)$ 
determinant expression of the $(2N \times L)$ partial domain 
wall partition function, while in this work, we start from 
the $(L\times L)$ determinant expression of the same object.

\subsection{The scalar product as a discrete KP $\tau$-function 
in the inhomogeneities}
In previous works \cite{foda.wheeler.zuparic, foda.schrader},
we showed that the scalar product is a discrete KP $\tau$-function
in the free rapidities, but we could 
not obtain the same result in terms of the inhomogeneities, 
even though the scalar product is symmetric in them.

As an application of the third determinant form of the scalar 
product obtained in this paper, we show that the scalar product 
is a discrete KP $\tau$-function in the inhomogeneities.

\subsection{Structure constants that are determinants at 
tree-level remain determinants at 1-loop level}
Given that the scalar product is a discrete KP $\tau$-function 
in the inhomogeneity variables $\{y\}$, it can be written in 
Casoratian form in any of the variables 
$y_i$, $i \in \{1, 2, \dots, L\}$. This allows us to show that 
the structure constants that can be expressed in determinant 
form \cite{Foda}, retain their determinant form in the presence
of 1-loop radiative corrections, when the latter are included 
along the lines proposed in 
\cite{gromov.vieira.short, 
      gromov.vieira.long}. This is an application 
of the results of this work to computations in weak-coupling 
$\cN = 4$ supersymmetric Yang-Mills.  

\section{Slavnov's scalar product}
\label{slavnov}

\subsection{Six-vertex model}

\begin{figure}
\begin{center}
\begin{minipage}{4.3in}
\setlength{\unitlength}{0.00038cm}
%(xwidth,ywidth)(-xpos,-ypos)
\begin{picture}(20000,17000)(-4500,-13000)
%
%%%Vertex a_{+}%%%
\path(-2000,2000)(2000,2000)
\put(-2300,2250){\tiny 1}
\put(2000,2250){\tiny 1}
%%%
\put(-3250,2000){\scriptsize{$x$}}
%\path(-4500,2000)(-3500,2000)
%\whiten\path(-4000,2250)(-4000,1750)(-3500,2000)(-4000,2250)
%%%
\path(0000,0000)(0000,4000)
\put(0000,-400){\tiny 1}
\put(0000,4250){\tiny 1}
%%%
\put(-1100,-3000){\scriptsize$a(x,y)$}
\put(0,-1250){\scriptsize{$y$}}
%\path(0,-2500)(0,-1500)
%\whiten\path(-250,-2000)(250,-2000)(0,-1500)(-250,-2000)
%
%%%Vertex b_{+}%%%
\path(8000,2000)(12000,2000)
\put(7700,2250){\tiny 1}
\put(12000,2250){\tiny 1}
%%%
\put(6750,2000){\scriptsize{$x$}}
%\path(5500,2000)(6500,2000)
%\whiten\path(6000,2250)(6000,1750)(6500,2000)(6000,2250)
%%%
\path(10000,0000)(10000,4000)
\put(10000,-400){\tiny 2}
\put(10000,4250){\tiny 2}
%%%
\put(8900,-3000){\scriptsize$b(x,y)$}
\put(10000,-1250){\scriptsize{$y$}}
%\path(10000,-2500)(10000,-1500)
%\whiten\path(9750,-2000)(10250,-2000)(10000,-1500)(9750,-2000)
%
%%%Vertex c_{+}%%%
\path(18000,2000)(22000,2000)
\put(17700,2250){\tiny 1}
\put(22000,2250){\tiny 2}
%%%
\put(16750,2000){\scriptsize{$x$}}
%\path(15500,2000)(16500,2000)
%\whiten\path(16000,2250)(16000,1750)(16500,2000)(16000,2250)
%%%
\path(20000,0000)(20000,4000)
\put(20000,-400){\tiny 2}
\put(20000,4250){\tiny 1}
%%%
\put(18900,-3000){\scriptsize$c(x,y)$}
\put(20000,-1250){\scriptsize{$y$}}
%\path(20000,-2500)(20000,-1500)
%\whiten\path(19750,-2000)(20250,-2000)(20000,-1500)(19750,-2000)
%
%%%Vertex a_{-}%%%
\path(-2000,-8000)(2000,-8000)
\put(-2300,-7750){\tiny 2}
\put(2000,-7750){\tiny 2}
%%%
\put(-3250,-8000){\scriptsize{$x$}}
%\path(-4500,-8000)(-3500,-8000)
%\whiten\path(-4000,-7750)(-4000,-8250)(-3500,-8000)(-4000,-7750)
%%%
\path(0000,-10000)(0000,-6000)
\put(0000,-10400){\tiny 2}
\put(0000,-5750){\tiny 2}
%%%
\put(-1100,-13000){\scriptsize$a(x,y)$}
\put(0,-11250){\scriptsize{$y$}}
%\path(0,-12500)(0,-11500)
%\whiten\path(-250,-12000)(250,-12000)(0,-11500)(-250,-12000)
%
%%%Vertex b_{-}%%%
\path(8000,-8000)(12000,-8000)
\put(7700,-7750){\tiny 2}
\put(12000,-7750){\tiny 2}
%%%
\put(6750,-8000){\scriptsize{$x$}}
%\path(5500,-8000)(6500,-8000)
%\whiten\path(6000,-7750)(6000,-8250)(6500,-8000)(6000,-7750)
%%%
\path(10000,-10000)(10000,-6000)
\put(10000,-10400){\tiny 1}
\put(10000,-5750){\tiny 1}
%%%
\put(8900,-13000){\scriptsize$b(x,y)$}
\put(10000,-11250){\scriptsize{$y$}}
%\path(10000,-12500)(10000,-11500)
%\whiten\path(9750,-12000)(10250,-12000)(10000,-11500)(9750,-12000)
%
%%%Vertex c_{-}%%%
\path(18000,-8000)(22000,-8000)
\put(17700,-7750){\tiny 2}
\put(22000,-7750){\tiny 1}
%%%
\put(16750,-8000){\scriptsize{$x$}}
%\path(15500,-8000)(16500,-8000)
%\whiten\path(16000,-7750)(16000,-8250)(16500,-8000)(16000,-7750)
%%%
\path(20000,-10000)(20000,-6000)
\put(20000,-10400){\tiny 1}
\put(20000,-5750){\tiny 2}
%%%
\put(18900,-13000){\scriptsize$c(x,y)$}
\put(20000,-11250){\scriptsize{$y$}}
%\path(20000,-12500)(20000,-11500)
%\whiten\path(19750,-12000)(20250,-12000)(20000,-11500)(19750,-12000)
%
\end{picture}
\end{minipage}
\end{center}

\caption{The six vertices and their associated Boltzmann weights.} 
\label{fig-6v}
\end{figure}
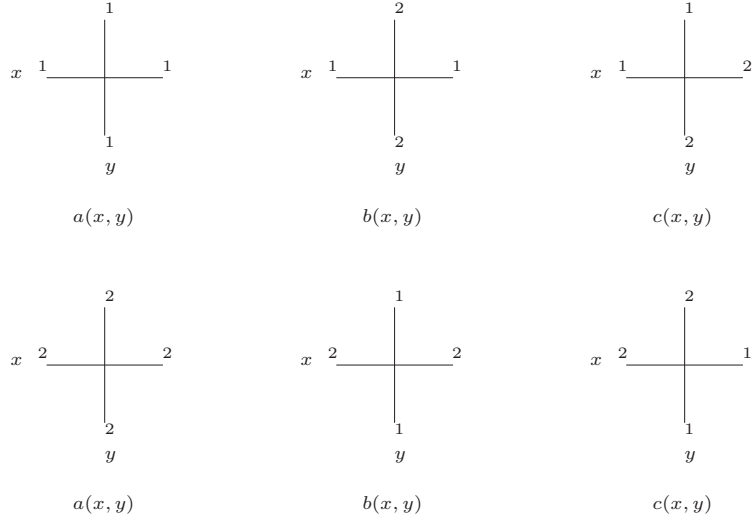

We consider the six-vertex model in the rational 
para\-metr\-iz\-ation, with the following  normalization of the 
Boltzmann weights
\begin{align}
a(x,y) = 1,
\quad\quad
b(x,y) = \frac{x-y}{x-y+1},
\quad\quad
c(x,y) = \frac{1}{x-y+1}
\label{rat-wt}
\end{align}
The assignment of weights to the vertices of the model is shown 
in Figure {\bf\ref{fig-6v}}.

\subsection{Scalar product}

Let $\{b\}_N = \{b_1,\dots,b_N\},
\{x\}_N = \{x_1,\dots,x_N\},
\{y\}_L =  \{y_1,\dots,y_L\}$ 
be three sets of rapidities\footnote{
To simplify the notation, we often leave the subscripts off 
the sets $\{b\}_N, \{x\}_N,\dots$ when the cardinality is clear from 
the context.}. Assume that their cardinalities satisfy 
$2N \leq L$ and that $\{b\}$ satisfy the Bethe equations
\begin{align}
\prod_{k=1}^{L}\ll
\frac{b_i-y_k+1}{b_i-y_k}
\rr
=
-
\prod_{j =1}^{N}\ll
\frac{b_i-b_j+1}{b_i-b_j-1}
\rr,
\quad
\forall\ 1 \leq i \leq N
\label{bethe}
\end{align}
We define the scalar product $S(\{x\},\{b\} | \{y\})$ to be the partition function of 
the lattice shown in Figure {\bf\ref{fig-sp}}. This definition is a completely consistent graphical representation of the algebraic form of the scalar product \cite{korepin.book}. The {\it first} determinant expression for the scalar product was found by Slavnov in \cite{slavnov}. Slavnov's expression is given by\footnote{Here, and in all subsequent determinants, we use Greek indices to label the rows and Latin indices to label the columns.}
\begin{figure}

\begin{center}
\begin{minipage}{4.3in}

\setlength{\unitlength}{0.00035cm}
\begin{picture}(40000,16500)(7000,6500)

%%% B(u) lines %%%%%

\path(14000,20000)(32000,20000) \put(12000,20000){\tiny $b_N$} 
\put(14000,20500){\tiny 1}  
\put(31500,20500){\tiny 2}

\path(14000,18000)(32000,18000) 
\put(14000,18500){\tiny 1}  
\put(31500,18500){\tiny 2}

\put(12000,17600){\tiny\vdots}

\path(14000,16000)(32000,16000)
\put(12000,16000){\tiny $b_1$}
\put(14000,16500){\tiny 1}  
\put(31500,16500){\tiny 2}

%%% C(u) lines %%%%%

\path(14000,14000)(32000,14000)
\put(12000,14000){\tiny $x_N$}
\put(14000,14500){\tiny 2}  
\put(31500,14500){\tiny 1}

\path(14000,12000)(32000,12000)
\put(14000,12500){\tiny 2} 
\put(31500,12500){\tiny 1}

\put(12000,11600){\tiny\vdots}

\path(14000,10000)(32000,10000) \put(12000,10000){\tiny $x_1$}
\put(14000,10500){\tiny 2} 
\put(31500,10500){\tiny 1}

%%% quantum lines %%%%%

\path(16000,8000)(16000,22000) \put(15500,6000){\tiny $y_1$}
\put(16000,7000){\tiny 1}
\put(16000,22500){\tiny 1} 

\path(18000,8000)(18000,22000)
\put(18000,7000){\tiny 1}
\put(18000,22500){\tiny 1} 
 
\path(20000,8000)(20000,22000)
\put(20000,7000){\tiny 1}
\put(20000,22500){\tiny 1} 

\put(22000,6000){\tiny \dots}
 
\path(22000,8000)(22000,22000)
\put(22000,7000){\tiny 1}
\put(22000,22500){\tiny 1} 
 
\path(24000,8000)(24000,22000) 
\put(24000,7000){\tiny 1}
\put(24000,22500){\tiny 1} 

\path(26000,8000)(26000,22000) 
\put(26000,7000){\tiny 1}
\put(26000,22500){\tiny 1} 

\path(28000,8000)(28000,22000) 
\put(28000,7000){\tiny 1}
\put(28000,22500){\tiny 1} 

\path(30000,8000)(30000,22000) \put(29500,6000){\tiny $y_L$}
\put(30000,7000){\tiny 1}
\put(30000,22500){\tiny 1}

\end{picture}

\end{minipage}
\end{center}

\caption{Lattice representation of $S(\{x\}_N,\{b\}_N | \{y\}_L)$. In the algebraic Bethe Ansatz setting, the top $N$ horizontal lines denote $B(b_i)$ operators, while the bottom $N$ horizontal lines denote $C(x_i)$ operators. The values 1 assigned to the top and bottom of the lattice denote the pseudo-vacuum states $|vac\rangle$ and $\langle vac|$, respectively.}
\label{fig-sp}

\end{figure}
\begin{multline}
S\ll
\{x\},\{b\} \Big| \{y\}
\rr
=
\Delta^{-1}\{x\} \Delta^{-1}\{-b\}
\times
\\
\det
\ll
\frac{1}{x_{\alpha}-b_j}
\ll
\prod_{k\not=j}^{N}
(b_k-x_{\alpha}-1)
\prod_{l=1}^{L}
\frac{(x_{\alpha}-y_l)}{(x_{\alpha}-y_l+1)}
-
\prod_{k\not=j}^{N}
(b_k-x_{\alpha}+1)
\rr
\rr_{\substack{1\leq \alpha \leq N \\ 1 \leq j \leq N}}
\label{slavnov-det}
\end{multline}
where we adopt the notation
\begin{align}
\Delta\{x\}
=
\prod_{1 \leq i< j \leq N}
(x_j - x_i),
\quad
\Delta\{-x\}
=
\prod_{1 \leq i< j \leq N}
(x_i - x_j)
\end{align}
for the Vandermonde in a set of $N$ variables $\{x\} = \{x_1,\dots,x_N\}$.

\section{A scalar product that depends on extra parameters}
\label{extended}

We consider a domain wall partition function on an $L \times L$ lattice, 
whose rapidities are the union of the three sets 
$\{t\}_{L-2N} = \{t_1,\dots,t_{L-2N}\}$,
$\{b\}_N = \{b_1,\dots,b_N\}$, 
$\{x\}_N = \{x_1,\dots,x_N\}$ and whose inhomogeneities are
$\{y\}_L = \{y_1,\dots,y_L\}$. We denote such a partition 
function by $Z(\{x\},\{b\},\{t\} | \{y\})$. Its graphical 
version is shown in Figure {\bf\ref{fig-dwpf}}.

\begin{figure}

\begin{center}
\begin{minipage}{4.3in}

\setlength{\unitlength}{0.00035cm}
\begin{picture}(40000,20500)(7000,6500)

%%% B(u) lines %%%%%

\path(14000,24000)(32000,24000) \put(11000,24000){\tiny $t_{L-2N}$} 
\put(14000,24500){\tiny 2}  
\put(31500,24500){\tiny 1}

\put(12000,22700){\tiny\vdots}

\path(14000,22000)(32000,22000) \put(12000,22000){\tiny $t_1$} 
\put(14000,22500){\tiny 2}  
\put(31500,22500){\tiny 1}

\path(14000,20000)(32000,20000) \put(12000,20000){\tiny $b_N$} 
\put(14000,20500){\tiny 2}  
\put(31500,20500){\tiny 1}

\put(12000,17600){\tiny\vdots}

\path(14000,18000)(32000,18000) 
\put(14000,18500){\tiny 2}  
\put(31500,18500){\tiny 1}

\path(14000,16000)(32000,16000)
\put(12000,16000){\tiny $b_1$}
\put(14000,16500){\tiny 2}  
\put(31500,16500){\tiny 1}

%%% C(u) lines %%%%%

\path(14000,14000)(32000,14000)
\put(12000,14000){\tiny $x_N$}
\put(14000,14500){\tiny 2}  
\put(31500,14500){\tiny 1}

\path(14000,12000)(32000,12000)
\put(14000,12500){\tiny 2} 
\put(31500,12500){\tiny 1}

\put(12000,11600){\tiny\vdots}

\path(14000,10000)(32000,10000) \put(12000,10000){\tiny $x_1$}
\put(14000,10500){\tiny 2} 
\put(31500,10500){\tiny 1}

%%% quantum lines %%%%%

\path(16000,8000)(16000,26000) \put(15500,6000){\tiny $y_1$}
\put(16000,7000){\tiny 1}
\put(16000,26500){\tiny 2} 

\path(18000,8000)(18000,26000)
\put(18000,7000){\tiny 1}
\put(18000,26500){\tiny 2} 
 
\path(20000,8000)(20000,26000)
\put(20000,7000){\tiny 1}
\put(20000,26500){\tiny 2} 
 
\path(22000,8000)(22000,26000)
\put(22000,7000){\tiny 1}
\put(22000,26500){\tiny 2} 

\put(22000,6000){\tiny \dots}
 
\path(24000,8000)(24000,26000) 
\put(24000,7000){\tiny 1}
\put(24000,26500){\tiny 2} 

\path(26000,8000)(26000,26000) 
\put(26000,7000){\tiny 1}
\put(26000,26500){\tiny 2} 

\path(28000,8000)(28000,26000) 
\put(28000,7000){\tiny 1}
\put(28000,26500){\tiny 2} 

\path(30000,8000)(30000,26000) \put(29500,6000){\tiny $y_L$}
\put(30000,7000){\tiny 1}
\put(30000,26500){\tiny 2}

\end{picture}

\end{minipage}
\end{center}

\caption{Lattice representation of $Z(\{x\}_N,\{b\}_N,\{t\}_{L-2N} | \{y\}_L)$. The lattice is $L \times L$, with domain wall boundary conditions. The horizontal lattice lines are split into three subsets, corresponding to the rapidities $\{x\}_N$, $\{b\}_N$ and $\{t\}_{L-2N}$. The extension parameters are the set $\{t\}_{L-2N}$.}
\label{fig-dwpf}

\end{figure}
Using Izergin's determinant formula for the DWPF \cite{izergin} (applied to 
the case where the rapidities are of mixed 
type\footnote{The usual form of Izergin's determinant, where 
the rapidities and inhomogeneities are labelled uniformly, as 
$\{x\} = \{x_1, \dots, x_N\}$ and $\{y\} = \{y_1, \dots, y_N\}$ 
respectively, is
\begin{align}
Z\ll \{x\} \Big| \{y\} \rr
=
\Delta^{-1}\{x\} \Delta^{-1}\{-y\}
\prod_{\alpha,j} (x_{\alpha}-y_j)
\det
\ll 
\frac{1}{(x_{\alpha} - y_j)(x_{\alpha} - y_j +1)} 
\rr_{\substack{1\leq \alpha \leq N \\ 1\leq j \leq N}}
\label{usual-izergin}
\end{align}}), we have
\begin{multline}
 Z \ll\{x\},\{b\},\{t\} \Big| \{y\} \rr
=
\Delta^{-1}\{x\} \Delta^{-1}\{b\} \Delta^{-1}\{t\} \Delta^{-1}\{-y\}
\times
\\
\frac{\displaystyle{
\prod_{\alpha,j} (x_{\alpha}-y_j) \prod_{\beta,j} (b_{\beta}-y_j) 
\prod_{\gamma,j} (t_{\gamma}-y_j)
}}
{\displaystyle{
\prod_{\alpha,\beta} (b_{\beta}-x_{\alpha}) 
\prod_{\alpha,\gamma} (t_{\gamma}-x_{\alpha}) 
\prod_{\beta,\gamma} (t_{\gamma}-b_{\beta})
}}
\det
\ll
\begin{array}{c}
 \frac{1}{(x_{\alpha}-y_j)(x_{\alpha}-y_j+1)}
\\
\\
\frac{1}{(b_{\beta}-y_j)(b_{\beta}-y_j+1)}
\\
\\
\frac{1}{(t_{\gamma}-y_j)(t_{\gamma}-y_j+1)}
\end{array}
\rr_{\substack{
1 \leq \alpha \leq N \\ 1 \leq \beta \leq N \\
1 \leq \gamma \leq L-2N \\ 1 \leq j \leq L
}}
\label{dwpf} 
\end{multline}
where indices in the products range over precisely the same values as in the determinant, namely $1\leq \alpha,\beta \leq N$, $1\leq \gamma \leq L-2N$, $1 \leq j \leq L$. We adopt this convention in all analogous formulae which follow. One of the purposes of this paper is to prove the following result. The proof will be given over the course of 
Sections {\bf\ref{third}}--{\bf\ref{second=first}}.
\begin{myLemma}
The domain wall partition function $Z(\{x\},\{b\},\{t\} | \{y\})$, 
given by (\ref{dwpf}), is an $(L-2N)$ parameter 
extension of Slavnov's scalar product $S(\{x\},\{b\} | \{y\})$, given by (\ref{slavnov-det}). 
The extension parameters are precisely the variables 
$\{t_1, \dots, t_{L-2N}\}$.
\label{lem1}
\end{myLemma}
  
\section{A third determinant expression for the scalar product}
\label{third}

\begin{figure}

\begin{center}
\begin{minipage}{4.3in}

\setlength{\unitlength}{0.00035cm}
\begin{picture}(40000,16500)(7000,6500)

%%% C(u) lines %%%%%

\path(14000,20000)(32000,20000) \put(12000,20000){\tiny $b_N$} 
\put(14000,20500){\tiny 2}  
\put(31500,20500){\tiny 1}

\path(14000,18000)(32000,18000) 
\put(14000,18500){\tiny 2}  
\put(31500,18500){\tiny 1}

\put(12000,17600){\tiny\vdots}

\path(14000,16000)(32000,16000)
\put(12000,16000){\tiny $b_1$}
\put(14000,16500){\tiny 2}  
\put(31500,16500){\tiny 1}

\path(14000,14000)(32000,14000)
\put(12000,14000){\tiny $x_N$}
\put(14000,14500){\tiny 2}  
\put(31500,14500){\tiny 1}

\path(14000,12000)(32000,12000)
\put(14000,12500){\tiny 2} 
\put(31500,12500){\tiny 1}

\put(12000,11600){\tiny\vdots}

\path(14000,10000)(32000,10000) \put(12000,10000){\tiny $x_1$}
\put(14000,10500){\tiny 2} 
\put(31500,10500){\tiny 1}

%%% quantum lines %%%%%

\path(16000,8000)(16000,22000) \put(15500,6000){\tiny $y_1$}
\put(16000,7000){\tiny 1}
\put(16000,21000){\circle*{200}} 

\path(18000,8000)(18000,22000)
\put(18000,7000){\tiny 1}
\put(18000,21000){\circle*{200}} 
 
\path(20000,8000)(20000,22000)
\put(20000,7000){\tiny 1}
\put(20000,21000){\circle*{200}} 
 
\path(22000,8000)(22000,22000)
\put(22000,7000){\tiny 1}
\put(22000,21000){\circle*{200}} 

\put(22000,6000){\tiny \dots}
 
\path(24000,8000)(24000,22000) 
\put(24000,7000){\tiny 1}
\put(24000,21000){\circle*{200}} 

\path(26000,8000)(26000,22000) 
\put(26000,7000){\tiny 1}
\put(26000,21000){\circle*{200}} 

\path(28000,8000)(28000,22000) 
\put(28000,7000){\tiny 1}
\put(28000,21000){\circle*{200}} 

\path(30000,8000)(30000,22000) \put(29500,6000){\tiny $y_L$}
\put(30000,7000){\tiny 1}
\put(30000,21000){\circle*{200}}

\end{picture}

\end{minipage}
\end{center}

\caption{Lattice representation of $Z(\{x\}_N,\{b\}_N | \{y\}_L)$. The top boundary segments are summed over both colours $\{1,2\}$, which is indicated by the dots placed on these segments. This $2N \times L$ lattice is obtained from the DWPF lattice by trivializing the top $L-2N$ lines, or in other words, by sending the extension parameters which live on those lines to infinity.}
\label{fig-pdw}

\end{figure}
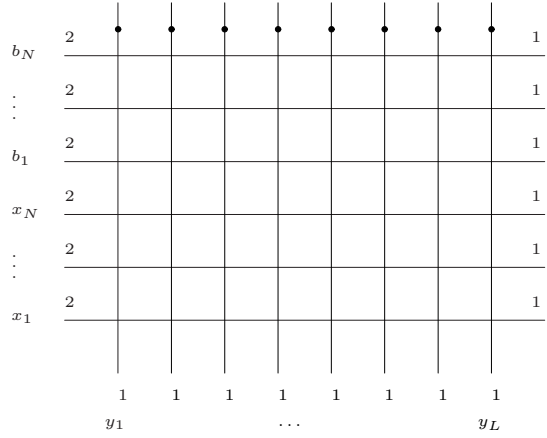

Define the function
\begin{align}
Z\ll
\{x\}, \{b\} \Big| \{y\}
\rr
=
\lim_{t_1,\dots,t_{L-2N} \rightarrow \infty}
\ll
t_1 \dots t_{L-2N}
Z\ll
\{x\}, \{b\}, \{t\} \Big| \{y\}
\rr
\rr
\label{limits}
\end{align}
This limit was studied in \cite{foda.wheeler.partial}, and the resulting 
object was called a \lq partial domain wall partition function\rq. This 
is in reference to the fact that  $Z(\{x\},\{b\}|\{y\})$ is the partition 
function of the rectangular lattice shown in Figure {\bf\ref{fig-pdw}}, with \lq partial domain wall boundary conditions\rq. 

Starting from the determinant expression (\ref{dwpf}), the limits 
in (\ref{limits}) can be taken explicitly. This procedure 
was explained in detail in \cite{foda.wheeler.partial}, so we only 
quote the result here,
\begin{multline}
 Z \ll \{x\},\{b\} \Big| \{y\} \rr
=
%\\
\frac{\displaystyle{
\prod_{\alpha,j} (x_{\alpha} - y_j) \prod_{\beta,j} (b_{\beta} - y_j)
}}
{\displaystyle{\Delta\{x\} \Delta\{b\} \Delta\{-y\}
\prod_{\alpha,\beta} (b_{\beta}-x_{\alpha})
}}
\det
\ll
\begin{array}{c}
 \frac{1}{(x_{\alpha}-y_j)(x_{\alpha}-y_j+1)}
\\
\\
\frac{1}{(b_{\beta}-y_j)(b_{\beta}-y_j+1)}
\\
\\
y_j^{L-2N-\gamma}
\end{array}
\rr_{\substack{
1 \leq \alpha \leq N \\ 1 \leq \beta \leq N \\
1 \leq \gamma \leq L-2N \\ 1 \leq j \leq L
}}
\label{pdwpf-ize} 
\end{multline}

To prove Lemma {\bf\ref{lem1}}, it is clearly sufficient to prove the following.
\begin{myLemma}
Let $S(\{x\},\{b\}|\{y\})$ and $Z(\{x\},\{b\}|\{y\})$ be Slavnov's scalar product (\ref{slavnov-det}) and the partial domain wall partition function (\ref{pdwpf-ize}), respectively. Assuming $\{b\}$ obey the Bethe equations, for all values of the variables $\{x\},\{y\}$ we have
\begin{align}
S\ll\{x\},\{b\} \Big| \{y\}\rr
=
(-)^N
Z \ll\{x\},\{b\} \Big| \{y\}\rr
\end{align}
\label{lem2}
\end{myLemma}
Lemma {\bf\ref{lem2}} is another version of the recent result of I Kostov and Y Matsuo \cite{kostov.matsuo}, who proved that the Slavnov scalar product is equal to a pDWPF. The result of \cite{kostov.matsuo} was in the context of the formula (\ref{pdwpf-kos}) for the pDWPF, which we discuss in the next section, whereas our result is in the context of equation (\ref{pdwpf-ize}) for the pDWPF. We call (\ref{pdwpf-kos}) and (\ref{pdwpf-ize}) the {\it second} and {\it third} expression for the scalar product, respectively.

\section{The third determinant expression equals the second} 
\label{third=second}

An alternative determinant expression for the pDWPF was found by Kostov 
in \cite{kostov.short.paper}. Kostov's expression for the pDWPF 
is\footnote{
Again, we remark that the usual form of Kostov's determinant 
is when the rapidities and inhomogeneities are labelled uniformly, as 
$\{x\} = \{x_1,\dots,x_N\}$ and $\{y\} = \{y_1,\dots,y_L\}$, respectively. 
In that case, it is given by
\begin{equation}
Z\ll \{x\}_N \Big| \{y\}_L \rr
=
\Delta^{-1}\{x\}
\det
\ll
x_{\alpha}^{j-1} - 
\prod_{l=1}^{L} 
\ll
\frac{x_{\alpha}-y_l}{x_{\alpha}-y_l+1} 
\rr
(x_{\alpha}+1)^{j-1}
\rr_{\substack{1\leq \alpha \leq N \\ 1 \leq j \leq N}}
\label{kostov-usual}
\end{equation}
In the case $N=L$, the determinant (\ref{kostov-usual}) becomes 
an alternative expression for the DWPF.
} 
\begin{multline}
 Z\ll\{x\},\{b\} \Big| \{y\}\rr
=
%\\
\frac{
\Delta^{-1}\{x\} \Delta^{-1}\{b\}
}
{\displaystyle{
\prod_{\alpha,\beta} (b_{\beta}-x_{\alpha})
}}
\det
\ll
\begin{array}{c}
x_{\alpha}^{j-1}- 
\prod_{l=1}^{L}
\ll
\frac{x_{\alpha}-y_l}{x_{\alpha}-y_l+1} 
\rr
(x_{\alpha}+1)^{j-1}
\\
\\
b_{\beta}^{j-1}- 
\prod_{l=1}^{L} 
\ll
\frac{b_{\beta}-y_l}{b_{\beta}-y_l+1} 
\rr
(b_{\beta}+1)^{j-1}
\end{array}
\rr_{\substack{
1 \leq \alpha \leq N \\ 1 \leq \beta \leq N \\
1 \leq j \leq 2N
}}
\label{pdwpf-kos} 
\end{multline}
In contrast to the determinant (\ref{pdwpf-ize}), which is 
$L \times L$, the determinant in (\ref{pdwpf-kos}) is $2N \times 2N$. 
A direct proof of the equivalence of the two determinants (\ref{pdwpf-ize}) 
and (\ref{pdwpf-kos}) was given in \cite{foda.wheeler.partial}. We will not 
repeat this proof here, and from now on treat (\ref{pdwpf-ize}) and 
(\ref{pdwpf-kos}) as interchangeable expressions for the pDWPF. 

\section{The second and third determinant expressions equal the first}
\label{second=first} 

In this section we give an alternative proof of Lemma
{\bf\ref{lem2}}. The original proof was given in
\cite{kostov.matsuo}. The basis of our proof is to define {\it
inhomogeneity\-/restricted} versions of both the scalar product
(\ref{slavnov-det}) and the pDWPF
(\ref{pdwpf-ize}),(\ref{pdwpf-kos}). The inhomogeneity\-/restricted 
scalar products were defined and calculated previously in 
\cite{kmt,wheeler.su2}.

\subsection{Inhomogeneity\-/restricted scalar products}
\label{sec-restrict-sp}

In this section, we consider variations of the size of the sets 
$\{x\},\{b\},\{y\}$. For that reason, it is necessary to restore 
subscripts to these sets to indicate their cardinality. For all 
$0\leq n \leq N$ we define
\begin{multline}
S\ll
\{x\}_n, \{b\}_N \Big| \{y\}_{N-n}, \{y\}^{N-n+1}_L
\rr
=
%\\
S\ll
\{x\}_N,\{b\}_N \Big| \{y\}_L
\rr
\Big|_{x_{N-i+1} = y_i,\ \forall\ 1 \leq i \leq N-n}
\label{restrict-sp}
\end{multline}
where $\{x\}_n     = \{X_1,\dots,x_n\}$, 
      $\{y\}_{N-n} = \{y_1,\dots,y_{N-n}\}$, 
      $\{y\}^{N-n+1}_{L} = \{y_{N-n+1},\dots,$ $y_L\}$. 
      We split up the dependence on the inhomogeneities deliberately, to indicate the separate symmetry in these two sets. The case $n=N$ is the scalar product (\ref{slavnov-det}) itself.

Starting from the formula (\ref{slavnov-det}), we can explicitly evaluate the function defined in (\ref{restrict-sp}). The result is the following hybrid determinant
\begin{multline}
S\ll
\{x\}_n, \{b\}_N \Big| \{y\}_{N-n}, \{y\}^{N-n+1}_L
\rr
=
\frac{\Delta^{-1}\{x\}_n \Delta^{-1}\{-b\}_N \Delta^{-1}\{-y\}_{N-n} }{
\displaystyle{\prod_{\alpha,\gamma} (y_{\gamma}-x_{\alpha})}}
\times
\\
\det
\ll
\begin{array}{c}
\frac{1}{x_{\alpha} - b_j}
\ll
\prod_{k\not=j}^{N}
(b_k-x_{\alpha}-1)
\prod_{l=1}^{L}
\frac{(x_{\alpha}-y_l)}{(x_{\alpha}-y_l+1)}
-
\prod_{k\not=j}^{N}
(b_k-x_{\alpha}+1)
\rr
\\
\\
\frac{1}{b_j-y_{\gamma}}
\prod_{k\not=j}^{N} (b_k-y_{\gamma}+1)
\end{array}
\rr_{\substack{1\leq \alpha \leq n \\ N-n \geq \gamma \geq 1 \\ 1 \leq j \leq N}}
\label{restrict-sp-det}
\end{multline}
Since this object comes from the first expression for the scalar
product (\ref{slavnov-det}), we call it the first expression for
the inhomogeneity\-/restricted scalar product.

\subsection{Properties of the inhomogeneity\-/restricted scalar product} 

Considering the inhomogeneity\-/restricted scalar 
product (\ref{restrict-sp-det}) as a function in $x_n$, we can show 
that it has the following properties.

{\bf A.} It is a meromorphic function in $x_n$ of the form
\begin{align}
S \ll \{x\}_n,\{b\}_N \Big| \{y\}_{N-n}, \{y\}^{N-n+1}_L\rr
=
\frac{P \ll \{x\}_n,\{b\}_N\Big|\{y\}_{N-n}, \{y\}^{N-n+1}_L \rr}
{
\prod_{i=1}^{n} \prod_{j=N-n+1}^{L} (x_i-y_j+1)
}
\label{sp-a}
\end{align}
where $P(\{x\}_n,\{b\}_N|\{y\}_{N-n}, \{y\}^{N-n+1}_L)$ is a polynomial 
of degree $L-N+n-1$ in $x_n$.

{\bf B.} It is symmetric in the set of variables $\{y_{N-n+1},\dots,y_L\}$.

{\bf C.} By setting $x_n = y_{N-n+1}$, we obtain the recursion relation
\begin{multline}
S \ll \{x\}_n,    \{b\}_N \Big| \{y\}_{N-n},   \{y\}^{N-n+1}_L \rr
\Big|_{x_n = y_{N-n+1}}
=
%\\
S \ll \{x\}_{n-1},\{b\}_N \Big| \{y\}_{N-n+1}, \{y\}^{N-n+2}_L \rr 
\label{sp-c}
\end{multline}

{\bf D.} In the case $n=0$, we have 
\begin{align}
S \ll \{x\}_0,\{b\}_N \Big| \{y\}_{N}, \{y\}^{N+1}_L \rr 
&=
\prod_{i,j=1}^{N}
\frac{(b_i-y_j+1)}{(b_i-y_j)}
Z \ll \{b\}_N \Big| \{y\}_N \rr
\label{sp-d}
\end{align}
where $Z(\{b\}_N|\{y\}_N)$ is the DWPF with rapidities 
$\{b\}_N = \{b_1,\dots,b_N\}$ and inhomogeneities 
$\{y\}_N = \{y_1,\dots,y_N\}$.

Since it is quite straightforward to verify that these statements are true, 
we only comment briefly on their proof. {\bf A} is proved by showing 
that all poles in $x_n$ in the denominator of (\ref{restrict-sp-det}) are 
cancelled by a zero resulting from setting two rows of the determinant equal. 
Similarly, one shows that the determinant itself has only poles at the points 
specified in the denominator of (\ref{sp-a}). Degree counting establishes the 
correct degree for the polynomial in the numerator.

{\bf B} is easy to prove, since the only place where 
$\{y_{N-n+1},\dots,y_L\}$ appear are in the factors 
$\prod_{l=1}^{L} (x_{\alpha}-y_l)/(x_{\alpha}-y_l+1)$, 
which are obviously symmetric with respect to these 
variables. {\bf C} follows from the definition (\ref{restrict-sp}) 
of the inhomogeneity\-/restricted scalar products.  
{\bf D} comes from comparing the $n=0$ case of (\ref{restrict-sp-det}) 
with Izergin's determinant representation of the DWPF (\ref{usual-izergin}).

Importantly, the properties {\bf A}--{\bf D} uniquely determine the set 
of scalar products (\ref{restrict-sp}). If another set of functions obey 
the same properties, they must be equal to the inhomogeneity\-/restricted 
scalar products.

\subsection{Inhomogeneity\-/restricted partial domain wall partition 
functions}

In analogy with Section {\bf\ref{sec-restrict-sp}}, we define
inhomogeneity\-/restricted versions of the pDWPF (\ref{pdwpf-ize}), 
(\ref{pdwpf-kos}). This is done in precisely the same way, namely, 
for all $0 \leq n \leq N$ we define

\begin{multline}
Z\ll
\{x\}_n, \{b\}_N \Big| \{y\}_{N-n}, \{y\}^{N-n+1}_L
\rr
=
%\\
Z\ll
\{x\}_N,\{b\}_N \Big| \{y\}_L
\rr
\Big|_{x_{N-i+1} = y_i,\ \forall\ 1 \leq i \leq N-n}
\label{restrict-pdwpf}
\end{multline}
The case $n=N$ is just the pDWPF itself. Explicit formulae 
for the inhomogeneity\-/restricted pDWPF can be obtained by 
starting from the determinants 
(\ref{pdwpf-ize}) or (\ref{pdwpf-kos}), and specializing 
the variables in the way prescribed by (\ref{restrict-pdwpf}). 
Doing this in the case of (\ref{pdwpf-ize}) gives
\begin{multline}
Z\ll
\{x\}_n, \{b\}_N \Big| \{y\}_{N-n}, \{y\}^{N-n+1}_L
\rr
=
\\
\frac{\displaystyle{
\prod_{\alpha,j}
%_{\alpha=1}^{n} \prod_{j=N-n+1}^{L}
(x_{\alpha}-y_j)
\prod_{\beta,j}
%_{\beta=1}^{N} \prod_{j=N-n+1}^{L}
(b_{\beta} - y_j)
}}
{\displaystyle{\Delta\{x\}_n \Delta\{b\}_N \Delta\{-y\}^{N-n+1}_L
\prod_{\alpha,\beta}
%_{\alpha=1}^{n} \prod_{\beta=1}^{N}
(b_{\beta}-x_{\alpha})
}}
\det
\ll
\begin{array}{c}
 \frac{1}{(x_{\alpha}-y_j)(x_{\alpha}-y_j+1)}
\\
\\
\frac{1}{(b_{\beta}-y_j)(b_{\beta}-y_j+1)}
\\
\\
y_j^{L-2N-{\gamma}}
\end{array}
\rr_{\substack{
1 \leq \alpha \leq n \\
1 \leq \beta \leq N \\
1 \leq \gamma \leq L-2N \\
N-n+1 \leq j \leq L
}}
\label{restrict-pdwpf-ize}
\end{multline}
where the indices of the products range over the same values 
as in the determinant, namely, 
$1\leq \alpha \leq n$, $1\leq \beta \leq N$, $N-n+1
\leq j \leq L$.\footnote{Interestingly, (\ref{restrict-pdwpf-ize}) does not depend on the inhomogeneities $\{y\}_{N-n} = \{y_1,\dots,y_{N-n}\}$, which we restrict to.} In the case of (\ref{pdwpf-kos}), we get
\begin{multline}
Z\ll
\{x\}_n, \{b\}_N \Big| \{y\}_{N-n}, \{y\}^{N-n+1}_L
\rr
=
\\
\frac{
\Delta^{-1}\{x\}_n \Delta^{-1}\{b\}_N \Delta^{-1}\{y\}_{N-n}
}
{
{\displaystyle{
\prod_{\alpha,\gamma}
(x_{\alpha}-y_{\gamma})
\prod_{\beta,\gamma}
(b_{\beta}-y_{\gamma})
\prod_{\alpha,\beta}
%_{\alpha=1}^{n} \prod_{\beta=1}^{N} 
(b_{\beta}-x_{\alpha})
}}
}
\det
\ll
\begin{array}{c}
y_{\gamma}^{j-1}
\\
\\
x_{\alpha}^{j-1} - 
\prod_{l=1}^{L}
\ll
\frac{x_{\alpha}-y_l}{x_{\alpha}-y_l+1} \rr
(x_{\alpha}+1)^{j-1}
\\
\\
b_{\beta}^{j-1} - 
\prod_{l=1}^{L} 
\ll
\frac{b_{\beta}-y_l}{b_{\beta}-y_l+1} \rr
(b_{\beta}+1)^{j-1}
\end{array}
\rr_{\substack{
1 \leq \gamma \leq N-n \\ 1 \leq \alpha \leq n \\ 1 \leq \beta \leq N \\ 1 \leq j \leq 2N 
}}
\label{restrict-pdwpf-kos}
\end{multline}
where we again remark that the indices in the products range over the values $1 \leq \alpha \leq n$, $1 \leq \beta \leq N$, $1\leq \gamma \leq N-n$. Ultimately, we will show that (\ref{restrict-pdwpf-kos}) and (\ref{restrict-pdwpf-ize}) are equal to the inhomogeneity\-/restricted scalar product (\ref{restrict-sp-det}). For that reason, hereafter we refer to them as the second and third expression for the inhomogeneity\-/restricted scalar product, respectively. 

\subsection{Properties of the inhomogeneity\-/restricted pDWPF}

Consider the determinant representations
(\ref{restrict-pdwpf-ize}) and (\ref{restrict-pdwpf-kos}) for the
inhomogeneity\-/restricted pDWPF, as a function in $x_n$. We claim
that the inhomogeneity\-/restricted pDWPF has the following properties.

{\bf A.} It is a meromorphic function in $x_n$ of the form
\begin{align}
Z \ll \{x\}_n, \{b\}_N \Big| \{y\}_{N-n}, \{y\}^{N-n+1}_L \rr
=
\frac{
P \ll \{x\}_n, \{b\}_N \Big| \{y\}_{N-n}, \{y\}^{N-n+1}_L \rr
}
{
\prod_{i=1}^{n} \prod_{j=N-n+1}^{L} (x_i-y_j+1)
}
\label{pdw-a}
\end{align}
where $P(\{x\}_n,\{b\}_N | \{y\}_{N-n}, \{y\}^{N-n+1}_L)$ is 
a polynomial of degree $(L-N+n-1)$ in $x_n$.

{\bf B.} It is symmetric in the set of variables $\{y_{N-n+1},\dots,y_L\}$.

{\bf C.} By setting $x_n = y_{N-n+1}$, we obtain the recursion 
relation
\begin{multline}
Z \ll \{x\}_n, \{b\}_N      \Big| \{y\}_{N-n}, \{y\}^{N-n+1}_L \rr
\Big|_{x_n = y_{N-n+1}}
=
%\\
Z \ll \{x\}_{n-1},\{b\}_{N} \Big| \{y\}_{N-n+1}, \{y\}^{N-n+2}_L \rr
\label{pdw-c}
\end{multline}

{\bf D.} In the case $n=0$, we have 
\begin{align}
Z \ll \{x\}_0,\{b\}_N \Big| \{y\}_{N}, \{y\}^{N+1}_L \rr 
&=
Z\ll \{b\}_N \Big| \{y\}^{N+1}_L \rr
\label{pdw-d}
\end{align}
where $Z(\{b\}_N | \{y\}^{N+1}_L)$ is the pDWPF with rapidities 
$\{b\}_N = \{b_1,\dots,b_N\}$ and inhomogeneities 
$\{y\}^{N+1}_L = \{y_{N+1},\dots,y_L\}$.

To prove these properties, it is convenient to freely change between the expressions (\ref{restrict-pdwpf-ize}) and (\ref{restrict-pdwpf-kos}). {\bf A} is proved using (\ref{restrict-pdwpf-kos}). One can easily check that all poles in $x_n$ in the denominator of (\ref{restrict-pdwpf-kos}) are cancelled by a zero from setting two rows of the determinant equal. The only poles which are present are the ones described by (\ref{pdw-a}), and degree counting establishes the correct degree for the polynomial in the numerator.

{\bf B} is proved using (\ref{restrict-pdwpf-ize}), which is invariant under simultaneously reordering the inhomogeneities in the Vandermonde $\Delta\{-y\}^{N-n+1}_L$ and those in the determinant. {\bf C} follows from the definition (\ref{restrict-pdwpf}) of the inhomogeneity\-/restricted pDWPF. {\bf D} is proved by considering the $n=0$ case of (\ref{restrict-pdwpf-ize}), when it is identically the pDWPF described in (\ref{pdw-d}). 

\subsection{Returning to proof of Lemma {\bf\ref{lem2}}}

The functions $S(\{x\}_n,\{b\}_N | \{y\}_{N-n}, $ $\{y\}^{N-n+1}_L)$ 
and $Z(\{x\}_n,\{b\}_N | \{y\}_{N-n}, \{y\}^{N-n+1}_L)$ satisfy the 
same set of properties {\bf A}--{\bf C}. The only apparent difference 
is their initial condition, property {\bf D}. In the following 
subsection we will show that because $\{b\}_N$ satisfy the Bethe 
equations, the right hand sides of equations (\ref{sp-d}) and 
(\ref{pdw-d}) are in fact equal up to the sign $(-)^N$. In doing so, 
we will have proved that 
\begin{multline}
S \ll \{x\}_n,\{b\}_N \Big| \{y\}_{N-n}, \{y\}^{N-n+1}_L \rr
=
(-)^N
Z \ll \{x\}_n,\{b\}_N \Big| \{y\}_{N-n}, \{y\}^{N-n+1}_L \rr 
\label{lem2-stronger}
\end{multline}
for all $0 \leq n \leq N$, due to the fact that the properties {\bf A}--{\bf D} are 
uniquely determining. This is of course sufficient to prove Lemma 
{\bf\ref{lem2}}, which corresponds to the case $n=N$ of equation 
(\ref{lem2-stronger}).

\subsection{Resolving the initial condition}

We begin by adjusting the right hand side of (\ref{sp-d}),
\begin{multline}
\prod_{i,j=1}^{N}
\frac{(b_i-y_j+1)}{(b_i-y_j)}
Z\ll \{b\}_N \Big| \{y\}_N \rr
=
\\
\Delta^{-1}\{b\}_N
\det
\ll
\prod_{k=1}^{N} 
\frac{(b_i-y_k+1)}{(b_i-y_k)}
b_i^{j-1} 
- 
(b_i+1)^{j-1} 
\rr_{1\leq i,j \leq N}
\end{multline}
which follows from Kostov's expression for 
the DWPF (see (\ref{kostov-usual}) with $N=L$). Hence the 
right hand side of (\ref{sp-d}) can be written as
\begin{multline}
\label{change-ic}
\prod_{i,j=1}^{N}
\frac{(b_i-y_j+1)}{(b_i-y_j)}
Z\ll \{b\}_N \Big| \{y\}_N \rr
=
\\
\Delta^{-1}\{b\}_N
\det
\ll
\prod_{k=N+1}^{L}
\frac{(b_i-y_k)}{(b_i-y_k+1)}
\prod_{k\not=i}^{N} 
\frac{(b_i-b_k+1)}{(b_i-b_k-1)}
b_i^{j-1} 
- 
(b_i+1)^{j-1} 
\rr_{1\leq i,j \leq N}
\end{multline}
where the last line follows from using the Bethe equations 
(\ref{bethe}) to modify every entry of the determinant. We denote 
the final determinant in (\ref{change-ic}) by 
\begin{align}
D_1(N,\kappa)
=
\det
\ll
\kappa_i
\prod_{k\not=i}^{N}
\frac{(b_i-b_k+1)}{(b_i-b_k-1)}
b_i^{j-1}
-
(b_i+1)^{j-1}
\rr_{1\leq i,j \leq N}
{\rm with}\ 
\kappa_i \equiv \prod_{k=N+1}^{L}
\frac{(b_i-y_k)}{(b_i-y_k+1)}
\end{align}
and from now on, treat it as a linear function in free 
variables $\{\kappa_1,\dots,\kappa_N\}$. On the other hand, using 
Kostov's pDWPF formula (\ref{kostov-usual}), the right hand side 
of (\ref{pdw-d}) is given by
\begin{multline}
Z \ll \{b\}_N \Big| \{y\}^{N+1}_L \rr
=
\Delta^{-1}\{b\}_N
\det
\ll 
b_i^{j-1} - \prod_{k=N+1}^{L} \frac{(b_i-y_k)}{(b_i-y_k+1)} (b_i+1)^{j-1} 
\rr_{1\leq i,j \leq N}
\label{2-ic}
\end{multline}
Up to the sign $(-)^N$, which it is necessary for us to introduce at some 
point, we write the determinant in (\ref{2-ic}) as
\begin{align}
D_2(N,\kappa)
=
\det
\ll
\kappa_i
(b_i+1)^{j-1}
-
b_i^{j-1}
\rr_{1\leq i,j \leq N}
\end{align}
where again 
$\kappa_i \equiv \prod_{k=N+1}^{L} \frac{(b_i-y_k)}{(b_i-y_k+1)}$, but we 
treat these as free variables. We prove that $D_1(N,\kappa) = D_2(N,\kappa)$.

$D_2(N,\kappa)$ is a linear function in $\kappa_N$. Evaluating it at 
$\kappa_N=0$, we obtain
\begin{multline}
D_2(N,\kappa)\Big|_{\kappa_N=0}
=
%\\
-
\left|
\begin{array}{cccc}
(\kappa_1 - 1) & (\kappa_1 \bb_1 - b_1) & \cdots & 
(\kappa_1 \bb_1^{N-1} - b_1^{N-1})
\\
\vdots & \vdots & & \vdots
\\
(\kappa_{N-1} - 1) & (\kappa_{N-1} \bb_{N-1} - b_{N-1}) 
& \cdots & (\kappa_{N-1} \bb_{N-1}^{N-1} - b_{N-1}^{N-1})
\\
\\
1 & b_N & \cdots & b_N^{N-1}
\end{array}
\right|
\end{multline}
where we have introduced the notation $\bar{b}_i = b_i+1$. Subtracting $({\rm column}\ j+1)/b_N$ from $({\rm column}\ j)$ for all $1 \leq j \leq N-1$, the final row of the determinant is only non-zero in the final entry. This reduces the size of the determinant by 1, and after extraction of common factors from the surviving rows, one obtains
\begin{align}
\label{rec1}
& D_2(N,\kappa)\Big|_{\kappa_N=0}
\\
&=
-
\prod_{i=1}^{N-1}(b_N-b_i)
\left|
\begin{array}{cccc}
(\tkappa_1 - 1) & (\tkappa_1 \bb_1 - b_1) & \cdots & (\tkappa_1 \bb_1^{N-2} - b_1^{N-2})
\\
\vdots & \vdots & & \vdots
\\
(\tkappa_{N-1} - 1) & (\tkappa_{N-1} \bb_{N-1} - b_{N-1}) 
& \cdots & (\tkappa_{N-1} \bb_{N-1}^{N-2} - b_{N-1}^{N-2})
\end{array}
\right|
\nonumber
\\
&=
- \prod_{i=1}^{N-1} (b_N-b_i)
D_2(N-1,\tkappa)
\nonumber
\end{align}
where we have defined $\tkappa_i \equiv \kappa_i \frac{(b_N-\bb_i)}{(b_N-b_i)}$. One can also evaluate the derivative with respect to $\kappa_N$,
\begin{align}
\partial_{\kappa_N} D_2(N,\kappa)
=
\left|
\begin{array}{cccc}
(\kappa_1 - 1) & (\kappa_1 \bb_1 - b_1) & \cdots & (\kappa_1 \bb_1^{N-1} - b_1^{N-1})
\\
\vdots & \vdots & & \vdots
\\
(\kappa_{N-1} - 1) & (\kappa_{N-1} \bb_{N-1} - b_{N-1}) 
& \cdots & (\kappa_{N-1} \bb_{N-1}^{N-1} - b_{N-1}^{N-1})
\\
\\
1 & \bb_N & \cdots & \bb_N^{N-1}
\end{array}
\right|
\end{align}
and subtracting $({\rm column}\ j + 1)/\bar{b}_N$ from $({\rm column}\ j)$ for all $1\leq j \leq N-1$, in analogy with above, we find that
\begin{align}
\label{rec2}
&
\partial_{\kappa_N} D_2(N,\kappa)
\\
&=
\prod_{i=1}^{N-1}(\bb_N-b_i)
\left|
\begin{array}{cccc}
(\kappa'_1 - 1) & (\kappa'_1 \bb_1 - b_1) & \cdots & (\kappa'_1 \bb_1^{N-2} - b_1^{N-2})
\\
\vdots & \vdots & & \vdots
\\
(\kappa'_{N-1} - 1) & (\kappa'_{N-1} \bb_{N-1} - b_{N-1}) 
& \cdots & (\kappa'_{N-1} \bb_{N-1}^{N-2} - b_{N-1}^{N-2})
\end{array}
\right|
\nonumber
\\
&=
\prod_{i=1}^{N-1}(\bb_N-b_i)
D_2(N-1,\kappa')
\nonumber
\end{align}
with the definition 
$\kappa_i' \equiv \kappa_i \frac{(\bb_N-\bb_i)}{(\bb_N-b_i)}$. Since (\ref{rec1}) 
and (\ref{rec2}) are recursion relations for the linear function 
$D_2(N,\kappa)$ at two different values of $\kappa_N$ (namely, $\kappa_N = 0$ and as $\kappa_N \rightarrow \infty$), together with the 
initial condition $D_2(1,\kappa) = \kappa_1 -1$, they determine it uniquely.

Following essentially the same procedure discussed above, but applied to 
the determinant $D_1(N,\kappa)$, one can similarly show that 
\begin{align}
D_1(N,\kappa) \Big|_{\kappa_N=0} &= 
-\prod_{i=1}^{N-1} (b_N-b_i) D_1(N-1,\tkappa)
\\
\partial_{\kappa_N} D_1(N,\kappa) &= 
\prod_{i=1}^{N-1} (\bb_N-b_i) D_1(N-1,\kappa')
\end{align}
Hence $D_1(N,\kappa)$ and $D_2(N,\kappa)$ satisfy the same two 
recursion relations, and since $D_1(1,\kappa) = \kappa_1 -1$, they share 
the same initial condition. Hence they are equal for all $N \geq 1$.

\section{One-loop $\cN \! = \! 4$ SYM structure constants as determinants}

\subsection{The Gromov-Vieira mapping}

In \cite{gromov.vieira.short, gromov.vieira.long}, Gromov and
Vieira define the following mapping on any function
$f(\theta_1,\dots,\theta_N)$ of the variables
$\{\theta_1,\dots,\theta_N\}$,

\begin{align}
f \mapsto
[ f ]_{\theta}
=
f
\Big|_{\theta_1,\dots,\theta_N \rightarrow 0}
+
\frac{g^2}{2}
\sum_{i=1}^{N}
(\partial_{\theta_i} - \partial_{\theta_{i+1}})^2
f
\
\Big|_{\theta_1,\dots,\theta_N \rightarrow 0}
+
O(g^4)
\label{theta}
\end{align}
where $\partial_{\theta_{N+1}} \equiv \partial_{\theta_1}$.
Note that the mapping is defined to $O(g^2)$ in some small
expansion parameter $g$.

\subsection{Complete symmetric functions and discrete derivatives}

Following \cite{macdonald.book}, the complete symmetric functions 
$h_i\{y\}$ in
the set of variables $\{y\} = \{y_1,\dots,y_L\}$ are defined as
coefficients in a generating series,
\begin{align}
 \sum_{i=0}^{\infty} h_i\{y\} z^i = \prod_{l=1}^{L} \frac{1}{1-y_l z}
\end{align}
We define a discrete derivative $\Delta_l$ which acts on the complete
symmetric functions as
\begin{align}
\Delta_l h_i\{y\}
=
\frac{h_i\{y\} - h_i\{\widehat{y}_l\}}{y_l}
=
h_{i-1}\{y\}
\label{disc-deriv}
\end{align}
where the subscript $l$ is used to denote the $l$-th element of the set
$\{y\}$, and $\widehat{y}_l$ denotes the omission of that variable from
the set.

\subsection{Casoratian determinants}

We define a Casoratian matrix $\Omega$ to be one whose entries
$\omega_{i,j}$ are symmetric with respect to a set of variables $\{y\} =
\{y_1,\dots,y_L\}$, and satisfy
\begin{align}
\omega_{i,j+1}\{y\} = \Delta_l \omega_{i,j}\{y\}
\label{casorati-cond}
\end{align}
where $\Delta_l$ is the discrete derivative with respect to any variable
$y_l \in \{y\}$. The determinant of a Casoratian matrix is a Casoratian
determinant, and is a discrete analogue of the Wronskian.

Using the definition (\ref{disc-deriv}) of the discrete derivative, it is
easy to see that the determinant
\begin{align}
|\Omega|
=
\det
\ll
\omega_{i,j}
\rr_{1\leq i,j \leq L}
=
\det
\ll
\sum_{k=1}^{M} c_{ik} h_{k-j}\{y\}
\rr_{1 \leq i,j \leq L}
\label{casorati-1}
\end{align}
satisfies (\ref{casorati-cond}), and is therefore Casoratian, for
arbitrary $M \geq L$ and coefficients $c_{ik}$, $1\leq i \leq L$, $1 \leq
k \leq M$. Using the Jacobi-Trudi identity for Schur functions
\cite{macdonald.book}, we can write (\ref{casorati-1}) equivalently 
as
\begin{align}
|\Omega|
=
\Delta^{-1}\{-y\}
\det
\ll 
\sum_{k=1}^{M} c_{ik} y_j^{k-1} \rr_{1\leq i,j \leq L}
=
\Delta^{-1}\{-y\}
\det
\ll 
\sum_{k=1}^{M} c_{jk} y_i^{k-1} 
\rr_{1\leq i,j \leq L}
\label{casorati-2}
\end{align}
where the final equality is just from matrix transposition. 
We will take (\ref{casorati-2}) as our generic form of a Casoratian 
determinant. Casoratian determinants are $\tau$-functions of the discrete KP hierarchy (see, for example, \cite{foda.schrader} for a more detailed exposition). All results stated in the sequel apply to any determinant 
of the form (\ref{casorati-2}), and all such determinants can also be viewed as discrete KP $\tau$-functions in the $\{y\}$ variables.

\subsection{Action of Gromov-Vieira mapping on Casoratian determinants}

In \cite{foda.wheeler.partial}, we studied the action of the GV mapping on
partial domain wall partition functions, which are Casoratian determinants
in their inhomogeneities $\{y\}_L$. However the procedure outlined in
\cite{foda.wheeler.partial} applies generally to any Casoratian
determinant of the form (\ref{casorati-1}) (or equivalently,
(\ref{casorati-2})). We briefly review these results here.

Let $[|\Omega|]_y$ denote the GV mapping of the determinant
(\ref{casorati-2}) (the transposed version, for notational convenience),
with respect to the variables $\{y_1,\dots,y_L\}$. We claim that
\begin{align}
\Big[ |\Omega| \Big]_y
=
\left|
\begin{array}{c}
c_{j,1} \\ \vdots \\ c_{j,L-2} \\ c_{j,L-1} + g^2 L c_{j,L+1} \\ c_{j,L} +
g^2 L c_{j,L+2}
\end{array}
\right|_{1\leq j \leq L}
+
O(g^4)
\label{GV-result}
\end{align}
We will not prove this equation here, since full details can be found in
\cite{foda.wheeler.partial}. Equation (\ref{GV-result}) says that, up to
higher order corrections in $g$, the determinant structure of Casoratian
determinants is preserved under the GV mapping (\ref{theta}).

\subsection{Restricted scalar products are Casoratian determinants}

Returning to equation (\ref{restrict-pdwpf-ize}) for the
inhomogeneity\-/restricted scalar product
$S( \{x\}_n, \{b\}_N | \{y\}_{N-n}, $ $ \{y\}^{N-n+1}_L )$, 
we renormalize it as follows,
\begin{multline}
\mathbb{S}\ll
\{x\}_n, \{b\}_N \Big| \{y\}_{N-n}, \{y\}^{N-n+1}_L
\rr
\equiv
\\
\prod_{\alpha=1}^{n} \prod_{j=N-n+1}^{L}
(x_{\alpha}-y_j+1)
\prod_{\beta=1}^{N} \prod_{j=N-n+1}^{L}
(b_{\beta}-y_j+1)
S\ll
\{x\}_n, \{b\}_N \Big| \{y\}_{N-n}, \{y\}^{N-n+1}_L
\rr
\end{multline}
so that $\mathbb{S}(\{x\}_n, \{b\}_N | \{y\}_{N-n}, \{y\}^{N-n+1}_L)$
is a polynomial in $\{y_{N-n+1},\dots,y_L\}$. For convenience, define the
combined set of rapidities $\{X\}_{n+N} = \{X_1,\dots,X_{n+N}\}$, where
\begin{align}
X_i = x_i,\ \forall\ 1 \leq i \leq n,\quad X_{n+i} = b_i,\ \forall\ 1 \leq
i \leq N
\end{align}
It is then straightforward to show that
\begin{multline}
\mathbb{S}
\ll
\{x\}_n, \{b\}_N \Big| \{y\}_{N-n}, \{y\}^{N-n+1}_L
\rr
=
\\
\Delta^{-1}\{X\}_{n+N}
\Delta^{-1}\{-y\}^{N-n+1}_{L}
\det
\ll 
\sum_{k=1}^{L+2n} c_{ik}\{X\} y_j^{k-1} 
\rr_{\substack{1 \leq i \leq L-N+n \\ N-n+1 \leq j \leq L}}
\label{casorati-rSP}
\end{multline}
where the coefficients $c_{ik}\{X\}$ are given by
\begin{multline}
c_{ik}\{X\}
=
\left\{
\begin{array}{ll}
e_{2n+2N-k-1} \ll \{-X,-\bar{X}\} \backslash \{-X_i,-\bar{X}_i\} \rr,
&
1 \leq i \leq n+N
\\
\\
e_{3n+N+L-k-i+1} \{-X,-\bar{X}\},
&
n+N+1 \leq i \leq L-N+n
\end{array}
\right.
\end{multline}
with $\bar{X}_i \equiv X_i+1$ and where $e_k\{-X,-\bar{X}\}$ and $e_k(\{-X,-\bar{X}\} \backslash \{-X_i, -\bar{X}_i\})$ denote elementary
symmetric functions \cite{macdonald.book}, given by the generating series
\begin{align}
\sum_{k=0}^{2n+2N} e_k\{-X,-\bar{X}\} z^k
=
\prod_{j=1}^{n+N} (1-X_j z) (1-\bar{X}_j z)
\\
\sum_{k=0}^{2n+2N-2} e_k\ll \{-X,-\bar{X}\} \backslash \{-X_i, -\bar{X}_i\} \rr z^k
=
\prod_{j\not=i}^{n+N} (1-X_j z) (1-\bar{X}_j z)
\end{align}
From equation (\ref{casorati-rSP}), we see that the (renormalized) 
inhomogeneity\-/restricted scalar products are Casoratian determinants 
in the set of inhomogeneities $\{y_{N-n+1},\dots,y_L\}$. Hence their 
image under the GV mapping in these variables is given by (\ref{GV-result}). 
We conclude that the determinant structure of tree-level SYM$_4$ structure 
constants between three non-BPS states (which are equal to 
inhomogeneity\-/restricted scalar products, \cite{Foda}) is preserved 
under 1-loop corrections. This extends our previous result 
\cite{foda.wheeler.partial} in the context of tree-level SYM$_4$ 
structure constants between two BPS and one non-BPS state
(which are equal to pDWPF's, \cite{GSV}). Previously, we were 
unable to obtain such a result starting from the expression 
(\ref{restrict-sp-det}) for the inhomogeneity\-/restricted 
scalar product, which is not manifestly a Casoratian determinant 
in the inhomogeneities.

\vfill
\newpage

\section{Remarks}

\1 Our results extend to the trigonometric six-vertex model and the 
corresponding XXZ spin-$\frac{1}{2}$ chain without obstruction.

\2 We note the different dimensions of the determinants in the 
three expressions for the inhomogeneity\-/restricted scalar product, 
equations (\ref{restrict-sp-det}), (\ref{restrict-pdwpf-kos}) and 
(\ref{restrict-pdwpf-ize}). 
The first expression (\ref{restrict-sp-det}) comes directly from 
Slavnov's determinant for the scalar product, and is $N \times N$. 
The second expression (\ref{restrict-pdwpf-kos}) comes from the pDWPF 
expression for the scalar product in \cite{kostov.matsuo}, which has 
$2N$ rapidities. Hence (\ref{restrict-pdwpf-kos}) is $2N \times 2N$. 
The third expression (\ref{restrict-pdwpf-ize}), which is new in this 
work, comes from a specialization of an $L \! \times \! L$ determinant, 
and is $(L-N+n) \! \times \! (L-N+n)$. Since the second and third 
expressions, (\ref{restrict-pdwpf-kos}) and (\ref{restrict-pdwpf-ize}), 
are larger in size than the first (\ref{restrict-sp-det}), it is 
unclear whether these will be computationally advantageous in 
studies of $\mathcal{N} = 4$ SYM structure constants.

\3 It is tempting to conjecture that Slavnov-type scalar products in 
spin chains based on higher-rank algebras can also be obtained from 
Bethe-restricted versions of the higher-rank domain wall partition 
functions (giving a parameter-extended scalar product), 
in the limit where the extension parameters are decoupled 
\cite{kostov.private.communication.seoul}. 

\4 We have no statistical mechanical interpretation for the extra 
parameters in the parameter-extended Slavnov scalar product. 
Hopefully, these parameters will act as regularization parameters 
in computations of physical objects, such as correlation functions, 
to be removed at the end of the computation. It may also be that 
the parameter-extended scalar product is an expectation value of 
an operator that is characterized by the extension parameters. 

\5 The parameter-extended scalar product of Section {\bf 3}
is a discrete KP $\tau$-function in the extra parameters, 
which play the role of an extra set of Miwa variables.

\6 We used the fact that the third determinant expression for the 
inhomogeneity\-/restricted scalar product is 
a discrete KP $\tau$-function, and therefore can be explicitly  
written as a Casoratian determinant, together with the results  
of N Gromov and P Vieira \cite{gromov.vieira.short, 
                               gromov.vieira.long}, to 
show that tree-level SYM$_4$ structure constants, with 
three non-BPS states that are known to be determinants \cite{Foda}, 
remain determinants in the presence of 1-loop corrections. 
The results of \cite{gromov.vieira.short, 
                     gromov.vieira.long} actually extend 
to 2-loop radiative corrections, and our computations can
be straightforwardly extended to 2-loops. 
In \cite{serban}, D Serban argued that the results of 
\cite{gromov.vieira.short, gromov.vieira.long} extend to all 
loops, at least in the limit where all three operators are 
represented by asymptotically long spin chain states.

\7 The second determinant expression of Kostov and Matsuo is 
expressed as an expectation value of free charged fermions in
\cite{kostov.matsuo}. From this expectation value, one can 
deduce that the second determinant expression is a discrete 
KP $\tau$-function in the inhomogeneities\footnote{See equation 
{\bf A.6} and {\bf A.7} in \cite{kostov.matsuo}. We thank 
I Kostov for pointing this out to us.}. Writing this 
fermionic expectation value explicitly as a Casoratian 
determinant, it should be possible to use it to obtain 
the result of Section {\bf 7} in this work. It should also be
possible to obtain the result directly from the first 
determinant expression of Slavnov, but we expect this 
to be tedious. Having different expressions for the same 
object, we expect that each should be easier to use  for 
different purposes.

\8 While writing the results reported in this work, we became aware 
of \cite{fairbault.schuricht}, where the rational Gaudin model was 
studied, and it was shown that the scalar product between a generic 
state and Bethe eigenstate can be expressed as a domain wall partition 
function. This observation was based on replacing the original $N$-magnon 
Bethe eigenstate (coming from the action of $N$ spin-lowering operators 
on the highest weight state) with an equivalent $(L-N)$-magnon eigenstate 
(coming from the action of $L-N$ spin-raising operators on the lowest 
weight state). This is a different approach to the one reported in this 
paper, where we have considered the equivalence of the Slavnov scalar 
product and domain wall partition function without needing to change 
the Bethe roots. Hence the results of \cite{fairbault.schuricht} are 
essentially unrelated to the results in this work.    

\vfill
\newpage

\section*{Acknowledgments}
We thank I Kostov, D Serban and F Smirnov for inspiring discussions
and communications that led directly to this work.  
OF wishes to thank Prof Ch Ahn and Prof Ch Rim for their excellent 
hospitality at EWHA Womans University and Songang University, Seoul,
Korea, where this work was started.
OF is supported by the ARC, Australia. 
MW is supported by the CNRS, France.

\end{document}